\documentclass{emulateapj}
\usepackage{amssymb}

\shorttitle{Modified Velocity Centroids}
\shortauthors{Lazarian \& Esquivel}

\begin{document}

\title{Statistics of Velocity from Spectral Data: 
Modified Velocity Centroids}

\author{A. Lazarian\altaffilmark{1} and A. Esquivel\altaffilmark{1}}
\altaffiltext{1}{Astronomy Department, University of Wisconsin-Madison, 475 N.Charter
St., Madison, WI 53706, USA. \\ lazarian@astro.wisc.edu; esquivel@astro.wisc.edu}

\begin{abstract}

We address the problem of studying interstellar turbulence using 
spectral line data. We find a criterion when the velocity centroids
may provide trustworthy velocity statistics. To enhance the scope
of centroids applications, 
we construct a measure that we term ``modified velocity centroids''
(MVCs) and derive an analytical
solution that relates the 2D spectra of the modified centroids with the 
underlying 3D velocity spectrum. We test our results
 using synthetic maps constructed with data obtained through
simulations of compressible magnetohydrodynamical (MHD) turbulence.
We show that the modified velocity
centroids (MVCs) are  complementary to the the Velocity Channel Analysis (VCA)
technique. Employed together, they make determining
of the velocity spectral index more reliable and for wider variety of
astrophysical situations.

\end{abstract}


\section{Introduction}

The interstellar medium (ISM) is turbulent, as known from observations of 
non-thermal cloud velocities  \citep{Lar92}, and also expected on theoretical 
grounds, because of the
very large Reynolds numbers (defined as the ratio of inertial to viscous forces).
Understanding turbulence is crucial for understanding of  many physical 
processes,
including energy injection into the ISM, non-photoelectric heating
of the ISM, star formation, propagation of cosmic rays, and heat transport. 
(see review by \citealt{VazOP00},\citealt{CL03a}), and references therein.

Obtaining properties of ISM turbulence from observations is a 
long-standing problem
(see review by \citealt{L99}, and references therein).
The statistics of the random velocity field is essential for describing the
turbulence. 
To obtain the velocity statistics,
so-called velocity centroids (\citealt{M58}) have been frequently attempted.
However, the separation of the velocity and density contributions to the 
velocity centroids has always
been a problem. 
Therefore the relation between the statistics of
velocity and velocity 
centroids is frequently claimed to be trustworthy only when the
density fluctuations are negligible (see \citealt*{VOS03}).

To remedy this situation a statistical technique termed Velocity Channel 
Analysis, or in short VCA, was developed in  \citet{LP00}.
It has been successfully used to obtain the spectra of
turbulence in \ion{H}{1}~ in Small Magellanic Cloud and the Milky Way 
 (see \citealt{SL01},\citealt{DDG00}). 
However, the VCA requires turbulence to be supersonic and to follow power
laws (see \citealt{ELPC03} ). 

This paper will identify
a technique that can be reliably used with
turbulence even if the turbulence is subsonic and/or does not obey a power-law.
The latter case can occur when, for instance, self-gravity modifies the turbulence
on small scales.
In pursuing this goal we introduce modified velocity centroids (MVCs) and 
derive an analytical relation between the statistics of MVCs and the
underlying statistics of the 3D turbulent velocity.

\section{Model: Turbulence in Magnetized Gas}

There has been substantial progress in understanding of compressible
magnetohydrodynamical (MHD) turbulence (see a review by \citealt{CL03a})
that will guide us in adopting the model for this paper. 

Numerical simulations in \citet{CL02,CL03b} revealed anisotropic
spectra of Alfv\'{e}n and slow waves, as well as 
an isotropic spectrum of the fast waves. The anisotropies are elongated
along the local direction of magnetic field. However, since this local direction 
changes from one place to another, the anisotropy in the system of 
reference related
to the mean magnetic field is rather modest, i.e. of the order of $\delta B/B$,
where $\delta B$ is the dispersion of the random magnetic field.
For typical ISM $\delta B\sim B$ \citep{S72} and the expected
anisotropy is modest.
Therefore it is possible to characterize the turbulence using isotropic statistics
(see testing in \citealt{ELPC03} and in \S 4).

Only z-components of the velocity are available through Doppler shift
measurements.
In what follows we shall assume that the velocity and density of the gas 
can be presented as sums of a mean value and a fluctuating part:
$V_z=v_0+v$, $\rho_{tot}=\rho_0+\rho$, where $v_0\equiv \langle V_z \rangle$,
$\rho_0 \equiv \langle \rho_{tot} \rangle$. The fluctuating components
 $v$ and $\rho$ satisfy the condition $\langle v \rangle=0$,
$\langle \rho\rangle=0$.
Henceforth we use $\langle ...\rangle$ to denote ensemble averaging.
The correlation function of the z-component of velocities at two 
points $\mathbf{x_1}$ and  $\mathbf{x_2}$, separated a distance $r$ 
(with $\mathbf{r}=\mathbf{x_1}-\mathbf{x_2}$), denoted by subscripts $1$ and $2$
($V_{z,1}=v_0+v_1$, $V_{z,2}=v_0+v_2$) is
$\langle v_1 v_2\rangle \equiv B_{zz}(\mathbf{r})$. And it
is related to the component of the 3D spectrum through 
a Fourier transform (see \citealt{MY75}):
\begin{equation}
B_{zz} (\mathbf{r})=\int {\rm e}^{i\mathbf{k \cdot r}} F_{zz} (\mathbf{k}) d\mathbf{k},
\label{corr-spect}
\end{equation}
where
\begin{equation}
F_{zz}(\mathbf{k})=\left(F_{LL}(k)-F_{NN}(k)\right)k_z^2/k^2 + F_{NN}(k)
\label{spect}
\end{equation}
is the projection of the spectral tensor, expressed through the transversal
$F_{NN}$ and longitudinal $F_{LL}$ (normal and parallel to
 $\mathbf{r}$, respectively) 3D spectra of the velocity field, which
are functions of the wavenumber amplitude $k$.

Similarly, the two point density statistics can be given by the structure function 
$\langle (\rho_1-\rho_2)^2 \rangle\equiv d(r)$.

\section{Modified Velocity Centroids (MVCs)}

In what follows we consider the  emissivity of our media proportional
to the first power of density\footnote{This is the case for the \ion{H}{1}  $21cm$ line.
For recombination radiation (such as H$\alpha$ for instance) the
emissivity is proportional to the square of the density, and
modifications of the present technique are necessary.}
and no absorption present.
In this case, the intensity, $I(\mathbf{X})\equiv \int \rho_sdV$, 
is proportional to the column density,
where $\rho_s$ is the density of
emitters in the Position-Position Velocity (PPV) space (see \citealt{LP00}).
That is, $I(\mathbf{X})=\int \rho_s dV=\int \alpha \rho_{tot} dz$, where,
$\rho_{tot}$ is the density of atoms, $\alpha$ is a proportionality
constant, $dz$ denotes integration along the line of sight, and $\mathbf{X}$ is
a two dimensional vector in the plane of the sky.

The structure function of MVCs consists of two terms. It will
be clear from the discussion below that it may be safe to use ordinary
centroids when the second term is much smaller than the first term.
In particular,
\begin{equation}
M(\mathbf{R})=\left\langle \left[S(\mathbf{X_1})-S(\mathbf{X_2})\right]^2 -
\langle V^2 \rangle \left[I(\mathbf{X_1})-
I(\mathbf{X_2})\right]^2 \right\rangle
\label{M}
\end{equation}
where $\mathbf{R}=\mathbf{X_1}-\mathbf{X_2}$, $S(\mathbf{X})$ is a
``unnormalized centroid'' of the z-component of velocity $V_z$, 
and $\langle V_z^2 \rangle=v_0^2+v_{turb}^2+v_{thermal}^2$. 
To simplify our notations we shall omit the $turb$ and $z$-subscripts (also
for eq.(\ref{M})).
We do not discuss the effect of $v_{thermal}$ separately because for the
technique presented here it acts in the same way as regular motion,
and one can formally introduce $v_0'^2=v_0^2+v_{thermal}^2$.
$S(\mathbf{X})\equiv \int_\mathbf{X} V \rho_s dV$,
that is, at a given position $\mathbf{X}$, integrating
along the line of sight\footnote{In terms of practical data handling, to minimize the 
effect of absorption at the centers of the lines, subtracting of
mean values of velocity $v_0$ in the expression for $S(\mathbf{X})$
and $\langle V^2 \rangle$ may be advantageous.
In terms of $S(\mathbf{X})$ the use of $\delta S(\mathbf{
X})\equiv S(\mathbf{X}) -\langle S(\mathbf{X}) \rangle$ may be preferable.}.

From an observational standpoint, the velocity dispersion $\langle V^2 \rangle$ can
be obtained using the second moment of the spectral lines:

\begin{equation}
\langle V^2 \rangle\equiv \langle \int_\mathbf{X} V^2 \rho_s dV\rangle/
\langle \int_\mathbf{X} \rho_s dV
\rangle .
\label{V2}
\end{equation}  

To express the 2D statistics of the modified centroids $M(\mathbf{R})$ through
the underlying 3D velocity and density statistics we shall have to make 
several elementary transformations. First of all, the difference of integrals
that enter the expression for the modified centroids should be presented as
a double integral. Then using an elementary identity 
$(a-b)(c-d)=1/2[(a-d)^2+(b-c)^2-(a-c)^2-(b-d)^2]$ it is possible to find
(see \citealt{L95})
\begin{equation}
\langle (S(\mathbf{X_1})-S(\mathbf{X_2}))^2\rangle =\alpha^2 \int\int dz_1 dz_2 [D(\mathbf{r})-D(|z_1-z_2|)],
\label{S2}
\end{equation}
where $D\equiv \langle (V_1\rho_{tot,1}-V_2\rho_{tot,2})^2 \rangle$. Using the 
Millionshikov  hypothesis
(see \citealt{MY75}) to relate the fourth moments of the fields with the
second moments, namely that
$\langle h_1 h_2 h_3 h_4 \rangle\approx \langle h_1 h_2 \rangle \langle h_3 h_4 \rangle
+\langle h_1 h_3 \rangle \langle h_2 h_4 \rangle+ 
 \langle h_1 h_4 \rangle \langle h_2 h_3 \rangle$ we get sufficient simplifications
of the expression:
\begin{equation}
D\approx \langle V^2 \rangle d(r) +\langle \rho^2_{tot}\rangle \langle (V_1-V_2)^2\rangle
-\frac{1}{2}\langle (V_1-V_2)^2\rangle d(r) + c(\mathbf{r}),
\label{D}
\end{equation}
where $c(\mathbf{r})$ denote the cross terms arising from correlations between
the fluctuations of the velocity and density 
\begin{equation}
 c(\mathbf{r})= 2\langle v_1 \rho_2 \rangle^2 - 
4 v_0\langle v_1 \rho_1 \rho_2 \rangle
- 4 \rho_0 \langle \rho_1 v_1 v_2\rangle.
\label{cr}
\end{equation}

The correlations between velocity and density fluctuations have been studied earlier
\citep{LPVP01,ELPC03} and found not to have a strong impact on the VCA. In particular,
the first term in  eq. (\ref{cr}) was directly obtained in \citet{ELPC03}, and
numerical integration of the corresponding fluctuations provides fluctuations
in the centroid value that does not exceed 8\% of the MVCs values at small $|R|$.
In this paper we will neglect the cross terms and test this 
assumption by
analyzing synthetic maps of compressible MHD simulations. A more detailed analysis,
with more sets of data, and the cross terms,  will be included in a forthcoming
paper.

Eq. (\ref{D}) depends on both density and velocity statistics.
The dominant contribution arising from density fluctuations is the first term
of eq. (\ref{D}). Integration of this term yields the second term in eq. (\ref{M}),
as result the expression for the MVCs is
\begin{equation}
M(\mathbf{R})\approx \alpha^2 \langle V^2 \rangle \langle \rho^2_{tot}\rangle
\int \int dz_1 dz_2 \left[b(\mathbf{r})-b(|z_1-z_2|)\right],
\label{M2}
\end{equation}
where 
\begin{equation}
 b(\mathbf{r})= \frac{\langle (V_1-V_2)^2\rangle}{\langle V^2 \rangle}
\left(1 -\frac{1}{2}\frac{d(r)}{\langle \rho^2_{tot}\rangle}\right).
\label{b}
\end{equation}
If $(1/2)(d(r)/\langle \rho_{tot}^2 \rangle) \ll 1$ then eq.(\ref{b})
recovers the statistics of the velocity field.
This is certainly the case if the turbulence is subsonic, where
the amplitude of the density fluctuations is small, or when the density spectrum is steep
(i.e. most of the power residing at large scales) given that we are interested in the
fluctuations at small $R$.\footnote{By {\it steep} spectrum  we mean
$P_{3D}\propto k^n$ with $n<-3$. If $n>-3$ the spectrum is {\it shallow}
(see \citealt{LP00}).} However, if the amplitude fluctuations of density at small
scales is large (i.e. shallow spectrum or highly supersonic turbulence),
the contribution could be important. If we omit the density term in eq.(\ref{b})
and take a 2D Fourier transform to resulting distribution of MVCs we have:
\begin{equation}
P_2(\mathbf{K})=\frac{1}{4\pi^2}\int d^2\mathbf{R}~{\rm e}^{-i\mathbf{K \cdot R}} M(\mathbf{R}),
\label{P}
\end{equation}
where we have used again the notation of using capital letters to distinguish
2D vectors from the 3D counterparts.
Substituting eqs. (\ref{M2}), (\ref{corr-spect}), and (\ref{spect})
into eq. (\ref{P}) one gets
\begin{equation}
P_2(K)\sim C_1 F_{NN}(K,0)+C_2\delta(K)
\label{final}
\end{equation}
where $C_1$ and $C_2$ are constants 
($C_1=4\pi \alpha^2 \langle \rho_{tot}^2 \rangle L $, where $L$ is the extent of the 
emitting region) and the second term provides only a $\delta$
function contribution at $K=0$.
Thus eq. (\ref{final}) provides a relation\footnote{Our result
is similar to that in \citet{KSHH93} who used normalized centroids
given by eq. (\ref{C}) but implicitly assumed that $\rho=const$
(see also \citealt{VOS03}).}
between the spectrum of the modified
centroids and the transversal spectrum of the 3D velocity field.
We should note that the relation in eq. (\ref{final}) does not depend on a
particular form (power-law is a common assumption) of the underlying spectra.
If the turbulent velocity field is mostly
solenoidal (see \citealt*{PWP98,MGOR96}), the isotropic spectrum $E(k)$ 
that is used for describing hydrodynamic turbulence 
is uniquely defined
through $F_{NN}(k)$, namely, $E(k)\approx 4\pi k^2 F_{NN}(k)$.

\section{Numerical Testing}

In what follows we will include the normalized centroids in their usual
form (see \citealt{MB94})
\begin{equation}
C(\mathbf{X})\equiv \int_\mathbf{X} V \rho_s dV/\int_\mathbf{X} \rho_s dV .
\label{C}
\end{equation}
Intuitively it is clear that the power spectrum of the centroids given by
eq. (\ref{C}) should provide a better fit to the power spectrum of the velocity
(the spectral index, not the amplitude) than that of $S(\mathbf{X})$.
Indeed, the contributions of density fluctuations are mitigated by
the division over the column density in eq. (\ref{C}).

The testing was done using the files obtained through simulations of
compressible MHD turbulence. The description of
the code and the numerical simulations can be found in 
\citet{CL02} and in \citet{ELPC03}.
We use the same data cube that we used in \citet{ELPC03}.
Namely, the dimension of the cube is $216^3$, and the Mach number $2.5$.
The result of the MHD simulations are velocity, density and magnetic
field data. We use density and velocity to produce spectral line
data cubes, the procedure is described in \citet{ELPC03}. 

We calculated $S(\mathbf{X})$, $C(\mathbf{X})$, and $I(\mathbf{X})$
from synthetic spectral line data cubes.
The spectra $P_2(\mathbf{K})$ were obtained
by applying a fast Fourier transform to the 2D distribution
of the centroids and column density values.
The power spectrum of the MVCs was calculated subtracting $\langle V^2 \rangle$ times
the column density power spectrum, from the power spectrum of $S(\mathbf{X})$
(see eq. \ref{M}).
For the original data cubes (with a steep density spectrum),
the differences among all three types
of centroids were marginal for the dynamical range studied,
although unnormalized centroids systematically 
show more deviations from the velocity power spectrum.
To enhance the effect of density, we made the density shallow
by changing the Fourier components of the density amplitudes,
but, in order
to preserve the density-velocity correlations, we kept 
the phase information (see  \citealt{ELPC03}).
The results are shown in Fig 1. It is clear that the three measures
provide a good fit at large scales, but at small scales normalized
and unnormalized centroids tend to be density dominated.
An additional advantage of using MVC over the conventional normalized centroids
is that MVC provides the correct amplitude of the spectrum,
while normalized centroids at best give the correct index (logarithmic slope).

\begin{figure}
\includegraphics[width=88mm]{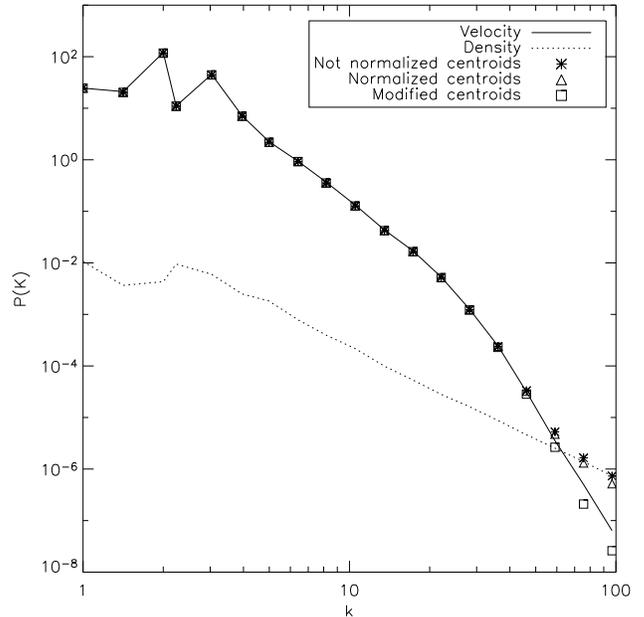}
\caption{The comparison of the 2D spectrum of the three types of
centroids with the underlying 3D spectrum of velocity
fluctuations. The solid line corresponds to the power spectrum of velocity 
for compressible MHD simulations with Mach number 
$\sim 2.5$. The dotted line corresponds to the spectrum of density
modified by changing the amplitudes the Fourier components of
the original spectrum. The modified density spectrum  $E(k)/k^2\sim k^{-2.5}$
for the inertial range. 
The vertical position of the normalized centroids 
({\it triangles}) is shifted vertically for visual purposes.
\label{fig:comp_mhd}}
\end{figure}

It is also clear that for a shallow density spectrum, the improvement
from unnormalized centroids to normalized centroids is marginal.

Our numerical study shows 
that MVCs are advantageous when the underlying
spectrum of density is dominated
by fluctuations at small scales.
And, the main advantage of MVCs is that
they allow an explicit statistical description of the procedures involved,
which makes the analysis more reliable. 

\section{Discussion}

Our study revealed that the MVCs can be successfully
used to obtain the velocity statistics from synthetic
observational data.
Another technique, namely, the VCA is complementary to the MVCs.
VCA is more
robust to density-velocity correlations. For instance, it was
shown in \citet{LP00} that VCA can provide correct
statistics of velocity fluctuations even in the case when the velocity
and density are correlated at their maximum given by the Cauchy-Swartz
inequality (see \citealt{MW70}). On the contrary,
if we assume such a maximum
correlation for the MVCs we will not get the correct scaling.
Comparing results obtained with the MVCs and the
VCA one can get a better handle on what are the velocity
statistics.

It is advantageous to combine VCA and MVCs techniques.
We can apply VCA to the largest scales,
where turbulence is supersonic, and study MVCs at all scaless, including
those where the turbulence becomes subsonic. 
The correspondence
of the spectral slope obtained with the two different techniques
would substantially increase the reliability of the result.
There is a more subtle, but important point. MVCs provide the
spectrum of solenoidal motions given by $F_{NN}$, while VCA is sensitive to both
potential and solenoidal motions. This may be used 
to get both potential and solenoidal motions in order to estimate the
role of compressibility. 
In addition, because
the thermal velocity acts in the VCA analysis the same way as the
thickness of the channel maps,  measurements of the spectrum with
the MVCs can provide a means of estimating the thermal velocity.
It is essential to combine the 
advantages of the techniques.

Our analysis of numerical
data  in \S 4 shows that the worries about the velocity centroids (see \citealt{S90}) 
although justified, may be somewhat exaggerated. 

If we {\it assume} that the criterion for the use
of the centroids is satisfied (that the second term in eq. (\ref{M})
is small compared to the first one) for earlier data sets,
the analysis of observational data in \citep{MB94} provides a range
of power-law indexes. Their results obtained with structure
functions if translated into spectra are consistent with
$E(k)=k^{\beta}$, where $\beta=-1.86$ with a standard
deviation of $0.3$. The Kolmogorov index $-5/3$
falls into the range of the measured values. L1228 exhibits
exactly the Kolmogorov index $-1.66$ as the mean value, 
while other low mass 
star forming regions L1551 and HH83 exhibit indexes close
to those of shocks, i.e. $\sim -2$. The giant molecular cloud regions show
shallow indexes in the range of $-1.9<\beta<-1.3$ (see \citealt*{MSB99}).
It worth noting that \citet{MB94}
obtained somewhat more shallow indexes that are closer to the
Kolmogorov value using autocorrelation functions. Those may
be closer to the truth as in the presence
of absorption in the center of lines, minimizing the regular velocity
used for individual centroids might make the results 
more reliable (see also first footnote in \S 3). 
Repeating of the study using MVCs and taking care of the absorption
effects may be advantageous.

The emergence of the Kolmogorov index for the compressible
magnetized gas may not be so surprising. 
\citet{CL02} for a range of Mach numbers reported
the existence of distinct Alfvenic, slow, and fast wave cascades.
The Alfv\'{e}n and slow waves result in Kolmogorov-type scaling.
The application of the
VCA to Small Magellanic Cloud \citep{SL01}
resulted in the velocity spectrum consistent with
those predictions. However, other researchers
(e.g. \citealt{VOS03}) reported indexes that
varied with the Mach number of the simulations. It is clear
that more  theoretical and observational work is necessary.

\section{Summary}

1. We derived a criterion when the velocity
centroids may reflect the actual underlying velocity statistics.
We introduced MVCs that may recover velocity statistics from
spectral line data in cases when the traditional centroid analysis fails.

2. Our numerical tests show that both the MVCs and the normalized
velocity centroids (see eq. (\ref{C})) successfully
recover the velocity information from data cubes obtained via
numerical simulations of the compressible MHD turbulence if the
density spectrum is steep. 

3. MVCs and the VCA are complementary independent techniques.
Obtaining the same power spectrum with both of them enhances confidence
in the result. MVCs are most useful when turbulence is subsonic
or/and does not follow the power law.
Additional information on the solenoidal versus
potential motions,
temperatures of gas etc. may be obtained combining these two techniques.
\\
\\
We thank the referee Anthony Minter for valuable suggestions that improved
this work.
The research by A.L. is supported by the NSF Grant AST-0125544.
A.E. acknowledges financial support from CONACyT (Mexico). 
Fruitful communications with Volker Ossenkopf are acknowledged.

\clearpage


\end{document}